\documentclass[journal,article,submit,moreauthors,pdftex,12pt]{article} 
\newcommand{\be}{\begin{eqnarray}}
\newcommand{\ee}{\end{eqnarray}}

\usepackage{cite}
\usepackage[percent]{overpic}
\usepackage{amsmath,amsthm,amssymb,latexsym,wasysym}

\title{Information-theoretic Neuro-correlates \\Boost Evolution of Cognitive Systems}
  
\author{Jory Schossau $^{1,2,\ast}$, Christoph Adami $^{2,3}$ and Arend Hintze $^{1,3,4}$}
\begin{document}
\maketitle

\begin{center}
$^{1}$ Department of Computer Science and Engineering\\
$^{2}$ BEACON Center for the Study of Evolution in Action\\
$^{3}$ Department of Microbiology \& Molecular Genetics\\
$^{4}$ Department of Integrative Biology\\
Michigan State University, East Lansing, MI 48824\\
$^{\ast}$ E-Mail: jory@msu.edu.
\end{center}

\begin{abstract}
Genetic Algorithms (GA) are a powerful set of tools for search and optimization that mimic the process of natural selection, and have been used successfully in a wide variety of problems, including evolving neural networks to solve cognitive tasks. Despite their success, GAs sometimes fail to locate the highest peaks of the fitness landscape, in particular if the landscape is rugged and contains multiple peaks. Reaching distant and higher peaks is difficult because valleys need to be crossed, in a process that (at least temporarily) runs against the fitness maximization objective. Here we propose and test a number of information-theoretic (as well as network-based) measures that can be used in conjunction with a fitness maximization objective (so-called ``neuro-correlates") to evolve neural controllers for two widely different tasks: a behavioral task that requires information integration, and a cognitive task that requires memory and logic. We find that judiciously chosen neuro-correlates can significantly aid GAs to find the highest peaks.
\end{abstract}


\section{Introduction} 

The last 50 years of research in Artificial Intelligence have taught us many things, but perhaps the most obvious lesson is that designing complex cognitive systems is extremely hard. Notwithstanding the success of chess-playing algorithms and self-driving cars, designing a brain that rivals the performance of even the smallest vertebrate has proven elusive. While the computational algorithms that are being deployed today on the aforementioned problems, (as well as on image classification via convolutional nets) are impressive, many researchers are convinced that none of these algorithms are cognitive in the sense that they display situational understanding. For example, the celebrated convolutional nets can easily be fooled~\cite{Nguyenetal2015} with trivial imagery, suggesting that they implement a sophisticated look-up table after all, with very little understanding. 

The failure of the design approach has been acknowledged by several groups of researchers that have chosen an entirely different approach, namely to use the power of evolution to create machine intelligence. This field of ``neuro-evolution''~\cite{yao1999evolving,floreano2008neuroevolution} is much less developed than the standard design approach, but it has made great strides in the last decade. It also has the advantage (compared to the design approach) that the approach is known to have resulted in human-level intelligence at least once. In the field of neuro-evolution, a Genetic Algorithm (GA)~\cite{Michalewicz1996} is used to evolve a program that, when executed, builds a ``neuro-controller''. This neuro-controller constitutes the brain of a simulated entity, which is called an agent.

Each program is evaluated via the performance of the agent, and programs that gave rise to successful brains are then replicated, and given proportionally more off-spring programs than unsuccessful programs. Because mutations are introduced in the replication phase, new types of programs are introduced every generation, trying out variations of the programs--and therefore variations of the brains. This algorithm, closely modeled on the Darwinian process that has given rise to all the biocomplexity on the planet today, has proven to be a powerful tool that can create neuro-controllers for a diverse set of tasks. 

Using evolution to create brains is no panacea, though. The use of Genetic Algorithms (GAs) to optimize performance (fitness) of behavior controllers is often hindered by the structure of complex fitness landscapes, that are typically rugged and contain multiple peaks. The GA, via its fitness maximization objective, will discover local peaks but may get stuck at sub-optimal peaks because crossing valleys is specifically {\em not} an objective. For a population to overcome the valleys in such a rugged landscape, programs must (at least temporarily) acquire deleterious mutations that are at odds with a simple reward system for optimization. This difficulty is typically overcome by increasing diversity in the population~\cite{DeJong1975}, by splitting the population into islands~\cite{Whitley1998,bitbol2014quantifying}, by using alternative objectives such as in novelty search~\cite{Lehman2008}, or by changing (and thus optimizing) the fitness function itself. Of these solutions, ``Diversity'' and ``Novelty'' can be computationally intensive, while fitness function optimization is very specific to every problem, and thus not a general solution.

Here we propose an alternative approach to fitness optimization in the evolution of cognitive controllers, that takes advantage of the insight that functioning brains have a certain number of characteristics that are a reflection of their network structure, as well as their information-processing capacity. If we were able to reward these features at the same time as rewarding the performance of the given task, it may be possible to evade the valleys of the landscape, and move on neutral ridges towards higher peaks. The idea of using multiple objectives in Genetic Algorithms is not at all new~\cite{Zhouetal2011}, and it has been used previously in neuro-evolution~\cite{Clune2013}. 

We present evidence from simulations of evolving virtual agents that establishes that it is possible to improve the performance of a GA, increase the rate of adaptation, and improve the performance of the final evolved solution, all by incorporating {\em neuro-correlates} of the evolving agent's brains into the fitness calculation. These neuro-correlates are metrics from cognitive science that attempt to measure \textit{``how well a brain is working''}, independently of the achieved fitness. These measures typically do not assess an agent's performance of a task because it is often difficult to relate task performance to cognitive ability. Ideally, these neuro-correlates either quantify the mode and manner that information is being processed, or in which manner the nodes of the network are connected. It is important that these neuro-correlates are agnostic of performance, as otherwise their reward would not open a new dimension in optimization.

The evaluation of an agent in any GA typically involves measuring the agent's performance for a given task or environment. We show that {\em multiplying} the performance of an agent with the value of its neuro-correlate will improve the speed of evolution and increase the ratio of perfect-performance evolved agents (a simple way of performing multi-objective optimization, see for example~\cite{deb2014multi}). This improvement can be traced back to an increase in the number of potentially beneficial mutations that may persist or sweep the population. If a mutation increases cognitive ability but does not yet have an effect on the agent's performance, then it is evolutionarily neutral and can be lost by drift. However, if the neuro-correlate shows an increase that is neutral with respect to performance, but improves cognition in some other form (and increases a neuro-correlate), then such a mutation is no longer neutral and is instead selected. In future generations, such an improvement might become beneficial for task performance. Therefore, using neuro-correlates in conjunction with performance allows these otherwise neutral mutations to stay in the population for longer or even promote them to fixation. Subsequent mutations have then a chance to take advantage of these changes that otherwise would have been lost to drift.

\section{Background}

We evolve agents to solve two very different tasks: a temporal-spatial integration task (active categorical perception, see~\cite{Beer1996,Beer2003,vandarteletal05,marstaller2013}) using embodied agents, and  a purely mathematical (disembodied) task that requires complex cognitive processing: the generation of random numbers using deterministic rules. The temporal-spatial integration task requires agents to observe and categorize blocks of different sizes falling toward them, by catching small blocks while avoiding large blocks. This task cannot be solved by a purely reactive machine (see~\cite{marstaller2013}) because agents must use memory in order to recognize the block's trajectory and predict where it will land. The task creates a fitness landscape known to be deceptive, as prior results have shown that only a small fraction (about 10\%) of populations result in an optimal solution. The sub-optimal agents usually get stuck on local peaks that deliver about 80\% of maximal fitness~\cite{marstaller2013}.

In the second set of experiments we investigate a task where agents are rewarded for generating long sequences of random numbers (without access to a random number generator or any other stochastic source). Agents are given an oscillating bit as an indicator of time, and assigned fitness based on the length of the dictionary generated from a Lempel-Ziv compression~\cite{welch1984technique} of the agent's output. Like the previous task, this task cannot be solved by a purely reactive machine, although would be trivially solved if the agents could access stochastic processes. However, because the agents use only deterministic processes we expect this task to require a great amount of temporal integration and memory in order for them to achieve a good amount of randomness. Indeed, generating random numbers is a known task to test cognitive ability and disability, in particular in the realm of Autism Spectrum Disorders and dementia~\cite{brugger1996random,baddeley1998random,jahanshahi2006random}. 

The standard substrate for neuro-evolution are Artificial Neural Networks (ANNs, see e.g.~\cite{RussellNorvig2003}), but we use here a different substrate (``Markov networks" or ``Markov Brains")  that has proven to be adept at behavioral decision-making tasks~\cite{albantakis2014evolution,edlund2011integrated,Olson2013SelfishHerd,chapman2013evolution,haley2014exploring,Olson150135,kvam2015,olson2013evolution}.
In contrast to ANNs in which neurons are continuous-valued and non-firing, neurons in Markov brains (MBs) are digital with only two states: quiescent or firing. Markov neurons can interact with any other neuron via arbitrary logical functions (as opposed to the ubiquitous transfer- and activation-function found in ANNs). We use MBs because we have experienced that they are computationally more powerful and more evolvable, while having a much smaller computational footprint than ANNs. ANNs on the other hand have a wide range of applications, and our results might generalize to those applications as well.

The logic functions that connect neurons, along with the topology of the network, are encoded directly by a string of bytes. The logic gates act on Markov variables (our equivalent of a neuron), and the output is written into other Markov variables. In a sense, the MB is defined by the edges between nodes, as it is the edges which carry all the computational power. Each gate is specified by a gene, and the beginning of each gene on the chromosome is specified by a particular combination of bytes--in our case, the ``start codon" (42,213). The bytes that follow determine the identity of the neurons it reads from, and the identifier of the neuron(s) it writes to. The subsequent bytes encode the logic of the gate, which can be done by simply encoding the truth table. While other MB implementations allow for stochastic logic gates, we confine ourselves to deterministic gates, which have a much more concise encoding (see Refs.~\cite{edlund2011integrated,chapman2013evolution,marstaller2013} for a more detailed description of MB encoding and function). 

There are alternative neural substrates that we could have studied here, including NEAT or hyperNEAT~\cite{Stanley2002}, genetic programming, or subsumption architecture machines~\cite{brooks1986robust}, etc.. These are all viable substrates for exploring the benefits of neuro-correlate aided evolution. In this contribution we focus on testing the general validity of the neuro-correlate augmented evolution approach. We do expect the results to depend on the underlying neural substrates, their evolvability, and how well each neuro-correlate can be assessed. In addition, our proposed method is easy to implement for other systems: A neuro-correlate must be measured and the resulting value multiplied by the associated performance. This should allow for a rapid testing of this method in other systems.

Despite evidence that an indirect encoding might be more advantageous~\cite{clune2011performance,d2011task,gauci2010indirect}, the direct encoding has been very successful in evolving controllers for virtual agents to solve a wide range of tasks~\cite{marstaller2010measuring,edlund2011integrated,chapman2013evolution,marstaller2013,albantakis2014evolution,hintze2014evolution}. In addition, these controllers have been instrumental in establishing some of the neuro-correlates used in the present study, which increases our confidence these measures perform as described. Next we describe the eight different neuro-correlates used to assess a controller's topology and performance. 

\newpage
\subsection{Network-theoretic neuro-correlates}

\subsubsection{Minimum-Description-Length}

The simplest neuro-correlate is, colloquially speaking, the largest possible brain size. It is difficult to define such a concept mathematically, but we can imagine that if we had a description of the brain in terms of the program that builds it, then the shortest such program would be the most concise description of the brain in a  Minimum Description Length (MDL) formalism, and larger MDLs could encode larger brains. The size of the genome that codes for our Markov brains could serve as a proxy for the brain MDL, but it is almost never the smallest description of the brain simply because the genome can add more ``empty tape'' instead of running out of space to encode more logic gates, for example using a gene duplication. Using the genome size (as proxy for MDL) as a neuro-correlate makes sense because it explicitly rewards genome expansion, rather than waiting for a fortuitous genome duplication to add the extra space. The genome size is directly proportional to the potential number of logic gates and thus the number of connections the agent phenotype might have, since the genome encodes the logic gates directly. Of course, under such a selective pressure genome length is almost guaranteed to be very different from the compression limit (the smallest program encoding the brain), but we can think of evolution as creating a selective pressure to compress the brain description as much as possible.

In our implementation of Markov Brains the genome size varies between $2,000$ and $20,000$ loci (each locus on the genome is a byte, so it can take values between 0 and 255) and can be affected by insertion- and deletion-mutations. We do not use the number of encoded logic gates to directly assess brain size for two reasons: First, each gate can have a different number of inputs and outputs, which influences the complexity of the gate. Second, gene duplications can create exact copies of a gene that codes for a gate, which changes the number of gates without (presumably) affecting the brain's function.

Because the connectivity of standard ANNs is fixed~\cite{RussellNorvig2003}, an MDL-like neuro-correlate does not exist there, but the Vapnik-Chervonenkis (VC)-dimension~\cite{VC1971} that bounds the learning capacity of a network could be a suitable alternative. Within more plastic systems ANNs that allow for encoding of connections such as NEAT~\cite{Stanley2002} the number of edges between neurons could be a proxy for potential brain size.

\subsubsection{Topological Complexity}

Brains are networks of neurons, and our MBs are networks of logic gates, both of which can be represented by graphs. Assessing the complexity of graphs is not a straight-forward task (see, e.g.,~\cite{McCabe1976}), but for the purpose of brain function some graph properties are obvious neuro-correlate candidates, and easy to measure. We first measure the {\em graph diameter} (GD) as the highest value in the distance matrix of the network - also known as the longest of all shortest paths between all node pairs.
The intuition behind using GD as a neuro-correlate is that information traveling along neurological pathways in brains with a large diameter has more time to interact with other neurons, in particular in other parts of the brain. If information takes longer to pass from sensors to actuators, it remains within the brain longer and therefore extends the agent's short-term memory. 

\subsubsection{Connectivity and Sparseness}
We measure the standard graph theoretic ``Gamma Index" (GI) or ``connectivity'' of the network as well as its converse, the network ``sparseness.'' The Gamma Index is the ratio of extant connections to all possible connections. For this measure multiple connections between nodes are treated like a single connection, otherwise there would be an infinite number of possible connections between all nodes of the network.

Current understanding of brain optimization and organization suggests that connections between neurons are costly~\cite{Ahn2006, Cherniak2004} and that this provides a strong selection pressure during evolution. Specifically, it has been shown that minimizing connections between neurons in a brain produces more modular and adaptable brains~\cite{Clune2013,huizinga2014evolving}. As you will see, this is not necessarily the case but depends on the task to evolve. Also, intuitively one might think that more connections are better, and thus optimizing for density might be as beneficial as for sparseness under the right circumstances. To incorporate this phenomenon, we use Sparseness and Gamma Index separately, as they reflect different aspects of brain connectivity. 

\subsubsection{Representations}
$R$, a measure for the amount of information that a brain represents within internal states, is a new information-theoretical measure of cognitive function that correlates with fitness~\cite{marstaller2010measuring,marstaller2013}, but is in principle separate from an agent's performance on a task. $R$ measures \textit{how much an agent knows about its environment above and beyond its current sensory information}. Representations can be thought of as internal models of the world that the agent can use to make decisions in conjunction with sensory signals, or even in the absence of sensory signals. Because of this, $R$ is often identified with ``what a brain knows about its world with its eyes closed".  
We can define $R$ as the information the mental states (described by a random variable $M$) of an agent have about the environment (described by random variable $E$), {\em given} its sensors (variable $S$)~\cite{marstaller2013}
\be
 R=I(E:M|S)=I(E:M)-I(E:M:S)\;. \label{rep}
\ee
In (\ref{rep}), $I(E:M|S)$ is the shared information between variables $E$ and $M$ given the sensor state $S$, $I(E:M)$ is the (unconditional) shared information between environment and brain internal states not including sensors), and $I(E:M:S)$ is the shared information between all three variables (a quantity that may become negative). The measure $R$ thus quantifies the correlation of mental states with world states, while subtracting the correlations coming from the sensors. This notion is not well-defined when using the correlation function, but it is when using information theory.   
$R$ increases in evolving agents when their tasks require knowledge or internal models about the environment~\cite{marstaller2013}. Agents without representations, or which are purely reactive agents, remain below optimal performance. Measuring these representations is simple as long as sensor, brain, and environmental states are directly observable (as they are here) causing only a small computational overhead. 

It is important to note that $R$ is not necessarily correlated with fitness. For instance, an agent might have representations about an incoming threat, but may not respond. Else, an agent may make decisions based solely on sensorial input, obtaining high fitness without representations. Therefore, representations do not necessarily make a prediction about an agent's performance, even though they are usually correlated. In addition, $R$ is not strictly speaking a neuro-correlate since it cannot be measured intrinsically. It is crucial that the correlate used does not allow predictions about performance, because otherwise the correlate in itself would be a proxy for fitness, and therefore optimizing a combination would not introduce a new element. $R$ satisfies this condition so we include this measure.

\subsubsection{Information Integration}
Agents solving a cognitively complex task must integrate information from different sensory inputs to come to a single decision. Sensory inputs must be compared to one another in the light of past experiences. The information-theoretic measure $\Phi$ is one way of quantifying a brain's information-integration capability~\cite{tononi2008consciousness,balduzzi2008integrated,balduzzi2009qualia,edlund2011integrated,albantakis2014evolution}. Unfortunately, computing $\Phi$ is computationally intensive, so much so that it is cumbersome on modern high-performance computers to calculate $\Phi$ exactly for brains of 16 neurons (for every agent at every update of an evolving population), and essentially infeasible for brains with more than 24 neurons\footnote{Because the computational complexity of $\Phi$ scales super-exponentially, calculating $\Phi$ for the brain of the lowly nematode {\it C. elegans} with 302 neurons, requires evaluating $\sim4.8\times 10^{457}$ partitions of the network, an absurd task.}.

Here we use the much more practical $\Phi_{\text{atomic}}$ which is a very good approximation of $\Phi$ at a much reduced computational cost~\cite{edlund2011integrated} ($\Phi_{\text{atomic}}$ has also been defined as "synergistic information"~\cite{barrett2011plos,ay2015entropy}). Specifically, $\Phi_{\text{atomic}}$ is given by the integrated information, but calculated for one specific partition of the network (while the standard $\Phi$ optimizes over partitions). The atomic partition is the one where each node is its own part, that is, the atomic partition segments the network into all nodes being individuals and not part of any other partition.
  
To define $\Phi_{\text{atomic}}$, we first define the information that is processed (in time) by the entire system. Let us define the network's state using the joint random variable $X = X^{(1)} X^{(2)} \dots X^{(n)}$, where $X^{(i)}$ represents the elements (nodes) of the system, where $X$ changes as time ($t$) progresses. Each variable $X_t$ is defined by a probability distribution $p(x_t)$ to find variable $X_t$ in state $x_t$. Each node $i$ progresses in time $X_0^{(i)} \to X_1^{(i)}$ and each $X_t^{(i)}$ is described by probability distribution $p(x_j^{i})$.

The information processed by the system from time step $t$ to $t+1$ is then given by
\be
I(X_t : X_{t+1}) =  \sum_{x_t,x_{t+1}} p(x_t,x_{t+1}) \text{log} \frac{p(x_t,x_{t+1})}{p(x_t)p(x_{t+1})}.
\ee
The measure $\Phi_{\text{atomic}}$ then quantifies how much of the information processed by the system cannot be explained by the sum of the information processed by each individual computational unit. Thus, in a sense  $\Phi_{\text{atomic}}$ quantifies how much processing is ``more than the sum of its parts", where the parts are defined by the individual neurons:
\be
\Phi_{\text{atomic}} = I(X_t : X_{t+1}) - \sum_{i=0}^{n}I(X_t^{(i)} : X_{t+1}^{(i)}) + \mathcal{I}. \label{phiatomic}
\ee
Here, $I(X_t^{(i)} : X_{t+1}^{(i)})$ is the information processed by the $i$th neuron, and $\mathcal{I}$ (called ``integration'' or ``multi-information'' in other work) measures the nonindependence between the network\linebreak variables~\cite{schneidman2003network,mcgill1954multivariate,tononi1994measure,lungarella2005methods,lungarella2006mapping} and is defined as
\be
\mathcal{I} = \sum_{i=0}^{n} H(X_t^{(i)}) - H(X_t). \label{multiinformation}
\ee

With both system information unexplained by part sums over time, and spatial integration measures of nonindependence, $\Phi_{\text{atomic}}$ represents both temporal and spatial network synergies.
Calculation of $\Phi_{\text{atomic}}$ then simplifies to
\be
\Phi_{\text{atomic}} = \sum_{i=0}^{n} H(X_t^{(i)} | X_{t+1}^{(i)}) - H(X_t | X_{t+1}) \label{eqPhiSimplified}
\ee

As with previous neuro-correlates, the act of integrating information does not imply that there will be an {\em action} upon such integrations. However, selecting agents with a higher $\Phi_{\text{atomic}}$ over others with the same performance guarantees the preservation of {\em potentially} beneficial mutations. In addition, we know that $\Phi_{\text{atomic}}$ is a limiting factor in the evolution of cognitive abilities: to perform a given task the agent requires a minimal amount of $\Phi_{\text{atomic}}$~\cite{Joshietal2013}, and a better performance necessitates a higher minimal amount of $\Phi_{\text{atomic}}$.
\begin{figure}
\begin{center}
  \includegraphics[width=0.7\columnwidth]{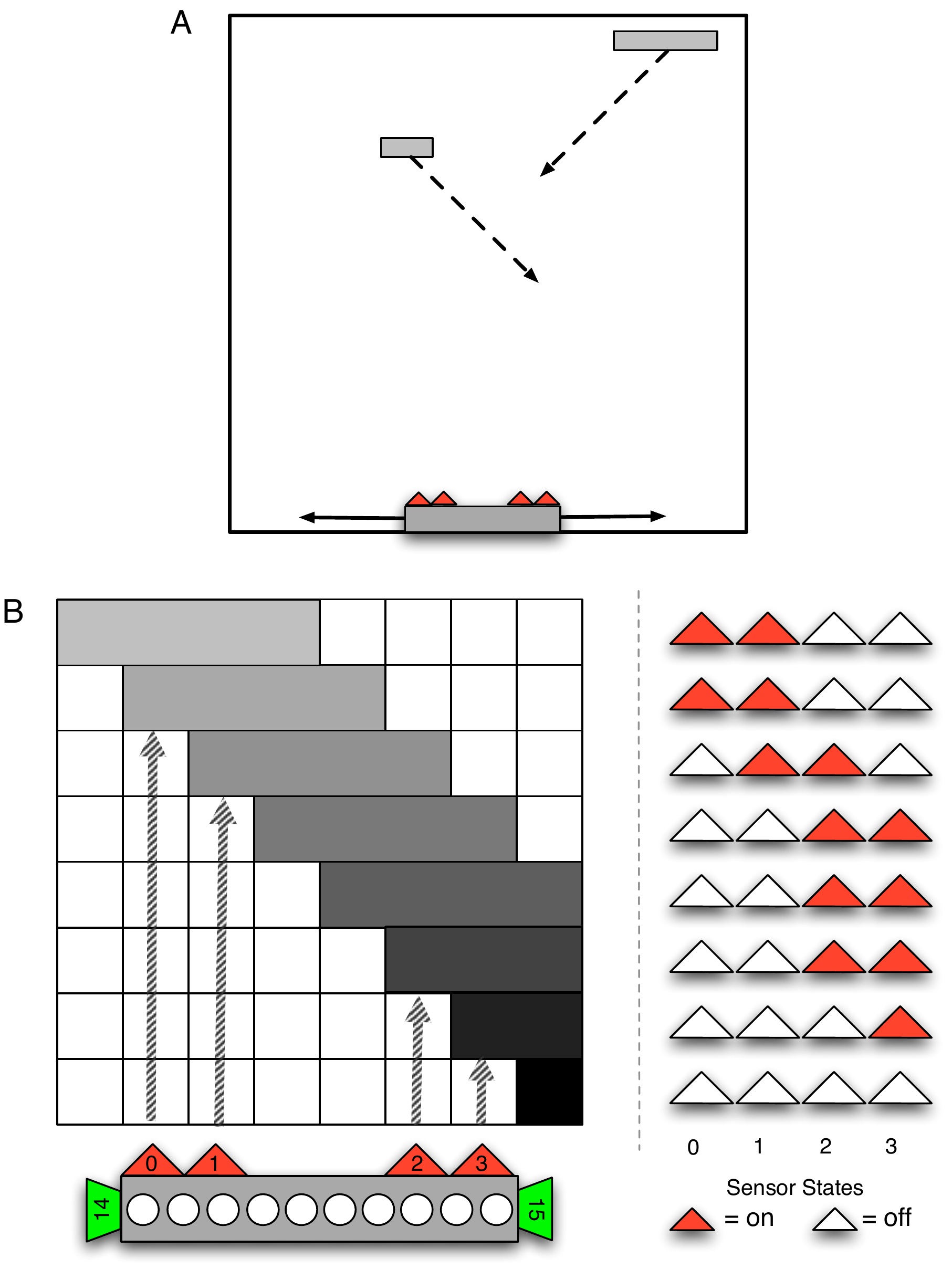} 
\end{center}
\caption{A: In the simulation, large or small blocks fall diagonally toward the bottom row of a $20\times20$ discrete world, with the agent on the bottom row. For the purpose of illustrating the task, a large brick is falling to the left, while a small brick is falling to the right. In simulations, only one block is falling at the time, and any one brick can fall either only to the left or only to the right. The agent is rewarded for catching small blocks and punished for catching large blocks. B:  A depiction of the agent's neurons (bottom left: triangles depict sensors, circles illustrate brain (internal) neurons, trapezoids denote actuators) and the sequence of activity patterns on the agent's 4-bit retina (right), as a large brick falls to the right. Reproduced from~\cite{marstaller2013}, with permission.
}
\label{fig:TaskIllustration}
\end{figure}

\subsubsection{Predictive Information}
Predictive information ($PI$) can be measured in several ways. It is the one-step mutual information a system has between time $t$ and $t+1$. MB animats have sensors and actuators, and $PI$ can be measured as a one-step mutual information of the sensors and future sensors, or the actuators and future actuators, or the sensors and future actuators, or the actuators and future sensors. Here we measure $PI$ of the sensors and future sensors ($PI_{\text{ss}}$), and sensors and future actuators ($PI_{\text{sa}}$)

\begin{equation}
\begin{aligned}
PI_{\text{ss}} &= I(S_{t} : S_{t+1})\\
PI_{\text{sa}} & = I(S_{t} : A_{t+1})
\end{aligned}
\end{equation}
where $S_{t}$ represents the sensor values at time $t$ and $A_{t}$ represents the actuator values at time $t$. 

An organism solving a physical task will move through the environment such that this information is increased---we typically do not look around randomly, but in a predictable manner. It has been shown that increasing $PI_{\text{ss}}$ can be advantageous for creating meaningful behavior on agents~\cite{ay2008predictive}.

Alternatively, predictive information can be understood as the information the sensors have about the actuators after the brain processed the sensor information. For a purely reactive agent, increasing this $PI_{\text{sa}}$ (for predictive information from sensor to motor) would be advantageous, because the actions of the agent become more appropriate to the action required. At the same time, if agents need to become less reactive but more dependent on their internal states $PI_{\text{sa}}$ should decrease after adaptation (as shown in ~\cite{edlund2011integrated}).

\subsection{Complex Environments}

We investigate the effect that rewarding neuro-correlates have on adaptation in two different environments. The first is a temporal-spatial integration task where an agent must catch or avoid blocks that fall towards it (see Fig:\ref{fig:TaskIllustration}). The task is an adaptation of the ``active categorical perception task" studied earlier~\cite{Beer1996,Beer2003,vandarteletal05}, and requires a comparison between past and present sensor inputs to make inferences about future optimal behavior. While the task is seemingly simple, the sensor modalities (embodiment) of the agent are limited in such a way that this becomes a complex problem to be solved by an agent~\cite{marstaller2013}. 

The second environment we use to optimize the agents is the generation of (pseudo) random numbers, and does not require embodiment of the brain. This task does not require any input, but the agent must produce a sequence of zeroes and ones with high entropy. This task is also used to assess cognitive abilities in humans. It is known that autism~\cite{rinehart2006pseudo}, schizophrenia~\cite{zlotowski1963behavioral}, as well as different forms or Alzheimer's disease can be diagnosed by analyzing a sequence of symbols generated by the human subject who was asked to produce a sequence that is as unpredictable as possible~\cite{rinehart2006pseudo,wagenaar1972generation,williams2002brief,brugger1996random}. This complex task involves memory~\cite{baddeley1998random}, processing~\cite{jahanshahi2006random}, and the ability to sequentially process information~\cite{brugger1996random} -- components that are also involved in the algorithmic generation of pseudo random numbers. It is unclear if an evolved Markov Brain random number generator resembles either a computer algorithm or the cognitive abilities found in humans. Nevertheless this task clearly qualifies as a complex problem requiring many components to work together, while at the same time it is not another example of an embodied agent. The randomness of the produced sequence is measured by its compressability using the Lempel-Ziv-Welch (LZW) algorithm~\cite{welch1984technique}.

One can think of many other complex environments for which this method of GA augmentation might be suitable, such as: navigation tasks~\cite{edlund2011integrated}, classification and perception tasks~\cite{chapman2013evolution}, or tasks that require an understanding of group behavior~\cite{olson2013predator,hintze2014evolution}. As long as neuro-correlates are measurable and doing so does not impose too high a computational overhead, this augmentation should be applicable. However, different environments could benefit differently from the neuro-correlates used--for example a one-layered perceptron in a classification task might not require internal representations. In such cases the representation measure $R$ might become useless.
Alternatively $\Phi_{\text{atomic}}$ might become meaningless in a task that does not require the integration of information


\section{Methods}

We performed evolutionary experiments to test how neuro-correlates affect the performance of a GA. 
(Our source code is available at https://gitlab.msu.edu/jory/entropy-2015-neuro-correlates). 
In each block-catching experiment the \textit{performance} of an agent was assessed by measuring the number of blocks correctly caught and correctly avoided in $80$ trials.
In the random-number-generation (RNG) experiment. the agents were given one bit that at each step changed from 0 to 1 and back for 500 updates. The output of the network over those 500 updates was collected and compressed. The number of symbols written into this compressed string was used as a proxy for the maximum entropy of that string. Highly regular or constant strings result in very short sequences after compression, while strings with higher entropy cannot be compressed that easily and result in longer strings. This environment has no particular state the world can be in, thus measuring $R$ is meaningless in this context.

Once performance is assessed, fitness can be calculated, upon which the GA's selection mechanism can operate. It is standard in this domain to use an exponential of performance ($1.1^{\text{performance}}$) as a fitness measure to encourage selection of more complex behavior, similar to~\cite{Lenski2003}. Using fitness alone is our control: the non-augmented case. We also explored the use of eight neuro-correlate measures for augmenting performance of the GA: minimum-description-length (MDL), diameter of the brain network (GD), amount of information integration ($\Phi_{\text{atomic}}$), amount of representation about the environment ($R$), Gamma Index (connectedness), sparseness, and two variations of predictive information: sensors $t$ to sensors $t+1$, and sensors $t$ to actuators $t+1$ ($PI_{\text{ss}}$ and $PI_{\text{sa}}$). Augmenting performance by a neuro-correlate is performed by a multiplication of the normalized neuro-correlate with the exponential performance of the agent.
Each evolutionary experiment is repeated 128 times and agents are allowed to adapt over $10,000$ generations (all evolutionary parameters are identical to~\cite{marstaller2013}, except duplication and deletion are identical at $0.05$). The population begins and ends at size $100$ and the mutation rate is $0.005$. At the end of each experiment the line of descent (LOD) is reconstructed~\cite{Lenski2003} and the results on the LOD are averaged over all replicates. The general form for fitness calculation used in this work is

\begin{equation} \label{eq:multiplexing}
\omega = 1.1^{\alpha} \prod \limits_{i=1}^N \left[\frac{\kappa_i}{\kappa_{i\text{max}}}+1\right]
\end{equation}
where $N$ is the set of neuro-correlates to use. These experiments used $N=1$ but the general form allows any number of neuro-correlates to augment performance together. $\alpha$ is the measure of performance, $\kappa$ the measure of the neuro-correlate, and $\kappa_{i\text{max}}$ the maximum theoretical value for the neuro-correlate used. While using $PI$ as a way to augment selection has previously used to minor success in the context of lifetime learning in embodied artificial intelligence~\cite{Zahedi2013} and attempted with linear additive function~\cite{Zahedi2013}, here we use a non-linear multiplicative function which emphasises the effect of both beneficial and deleterious mutations.

Alternatively, the distribution of fitness or neuro-correlates at the end of evolutionary adaptation is measured. The violin plots we use aggregate the replicate experiments and visualize the distribution. The final population contains genetic variation not yet culled by selection. To reduce such variation, we take from each experiment the organism on the line of descent three generations prior to the final generation. 


\section{Results and Discussion}

\begin{figure}
\begin{center}
\begin{overpic}[]{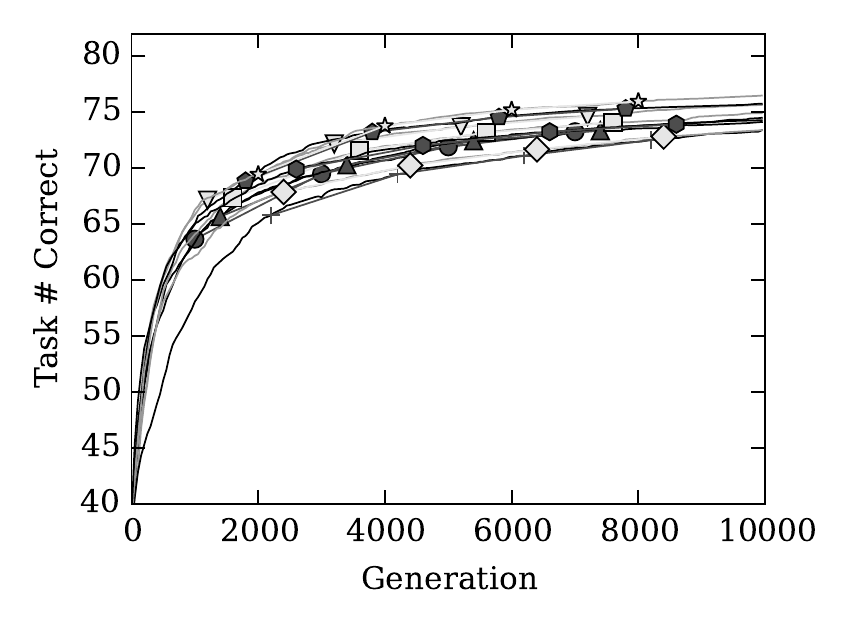}
 \put (11,5) {A}
\end{overpic}
\begin{overpic}[]{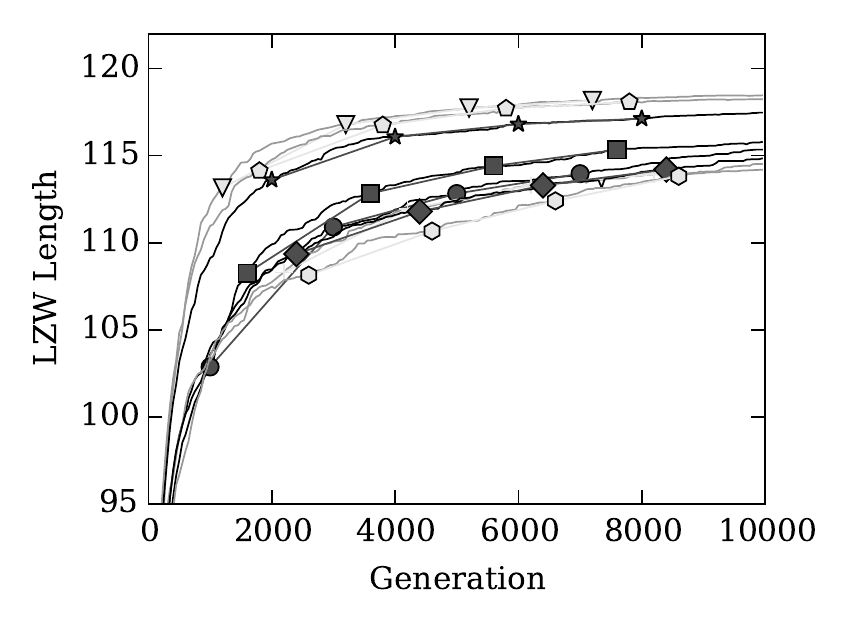}
 \put (12,5) {B}
\end{overpic}
\end{center}
\caption{A: Average of fitness over $10,000$ generations (over 128 replicates) for the block catching task. B: Average of fitness of $10,000$ generations (over 128 replicates) for generating random numbers. $\bullet$: the control using only performance in the GA; The effect of augmenting the fitness function (Eq:\ref{eq:multiplexing}) with $\Phi_{\text{atomic}}$ ($\triangledown$), with $R$ ($\blacktriangle$), with graph diameter ($\square$) on task performance, with minimum-description-length ($\pentagon$), with Gamma Index ($\bigstar$), with sparseness ($+$), with $PI_{\text{ss}}$ ($\diamond$), and with $PI_{\text{sa}}$ ($\hexagon$) on task performance.}
\label{fig:LOD}
\end{figure}

Results show that five of the eight proposed neuro-correlates improve adaptation when used to augment the GA in the block-catching task. The agent populations not only adapt faster, but also evolve to a higher average performance after $10,000$ generations (see Fig:\ref{fig:LOD} left). 
The highest fitness and most speedup of evolution is achieved by using potential brain size and sparseness, while using the Gamma Index performs worse than using performance alone. This paints a very interesting picture, as it seems to imply that pruning connections is better than having too many. However, increasing the potential brain size (which would add more connections) is the best accelerator of evolution. This suggests that new connections are important, but at the same time, connections that were added unnecessarily or are interfering should be removed quickly. All the other neuro-correlates ($\Phi_{\text{atomic}}$, diameter, and $R$) also improve the GA when optimizing agents to perform this task.

The results differ slightly for the RNG task (see Fig:\ref{fig:LOD} right). In general, the results for minimum-description-length, $\Phi_{\text{atomic}}$, and topology stay the same. However, the effects of sparseness and Gamma Index differ from the embodied task. For RNG, sparseness results in an inferior performance of the GA while Gamma Index promotes it. This suggests that in this task new connections are needed, but instead of being pruned, they need to be maintained, which is the opposite to what was observed in the block catching task. $R$ cannot be used in the context of RNG, and is therefore missing.

Predictive information is maximized in cases where reactive behavior is rewarded. In tasks that requires memory, maximizing predictive information can be detrimental (and is not the best predictor of fitness, see~\cite{edlund2011integrated}). It is possible that a predictive information with a larger time delay, or a conditional predictive information such as $I(S_t:A_t|A_{t-1})$ \cite{rivoire2011value} could produce better results. We plan on exploring those in the future.

Multiple parameter optimization (MPO) often leads to a whole host of problems, all well-described in the multiple-parameter optimization literature (for an overview see~\cite{deb2014multi}). In most of these problem cases, the parameters to be optimized work against each other in the form of trade-offs (one parameter can only be optimized at a cost of another). We observe this effect with Gamma Index and sparseness depending on the task to evolve, while all other neuro-correlates work synergistically with performance. See supplementary materials Fig:1-3 for evolutionary history interactions between neuro-correlates. We find that some neuro-correlates affect performance or each other antagonistically. While in our experiments this trade-off reduces the final performance of the evolved agents, it could be overcome using MPO. An objective which is antagonistic using our fitness function could be beneficial in MPO, which should be explored in the future.

\begin{figure}
\begin{center}
  \includegraphics[width=0.6\columnwidth]{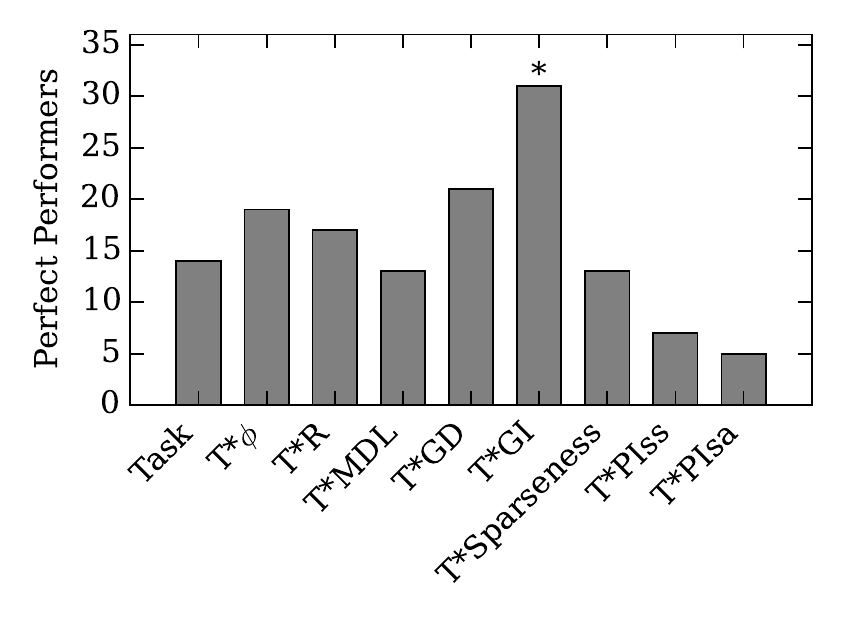} 
\end{center}
\caption{Absolute number of perfectly performing agents after $10,000$ generations evolved under the five different experimental conditions shown. * indicates significance under multiple hypothesis testing $p < 0.0001$.}
\label{fig:RatioPerfect}
\end{figure}

\subsection{Augmented selection and its effect on other neuro-correlates}

Using a neuro-correlate to augment a genetic algorithm can shorten the runtime requirements and may improve the overall performance of the evolved population depending on the neuro-correlate and objective. One might ask how augmenting selection using one neuro-correlate affects the other correlates not under direct selection.
Intuitively one would expect that selection for a particular neuro-correlate, in conjunction with performance, should increase not only performance (as discussed above) but also the measure itself. 
Similarly, since neither of the Predictive Information measures augment selection, then we expect no increase in Predictive Information when using them in conjunction with performance.  However, we find $PI_{\text{sa}}$ to increase when selecting for it together with performance in the RNG environment. 

\begin{figure}
\begin{center}
  \includegraphics[width=0.5\columnwidth]{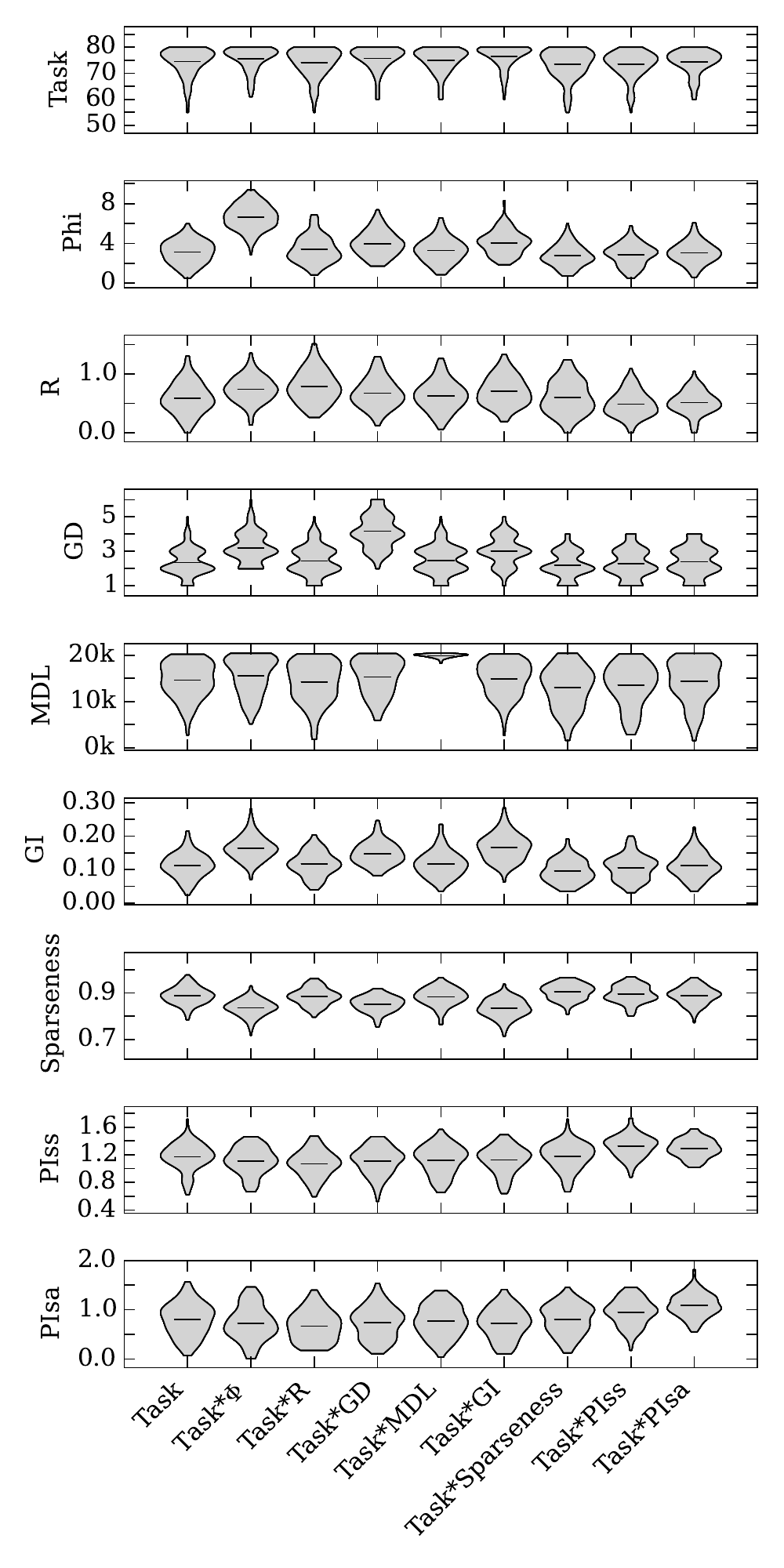} 
\end{center}
\caption{Comparison of different optimization methods at the end of evolution of the block catching task. The top row shows the performance distribution (gray) and average performance given the different selection regimes (MDL abbreviates minimum-description-length, which is the potential brain size based on the length of the encoding genome, GI abbreviates Gamma Index, GD graph diameter, and $PI_{\text{ss}}$ and $PI_{\text{sa}}$ are predictive information sensor $t$ to sensor $t+1$, and sensor $t$ to actuator $t+1$, respectively). Subsequent rows show the effect of the different selection regimes ($x$ axis) on the respective neuro-correlates.}
\label{fig:blockDistributionWithTask}
\end{figure}

Additionally we find no other effect of one neuro-correlate driving the evolution of another. The most prominent example of this effect is with Gamma Index driving the evolution of nearly all other neuro-correlates and vice versa (see Fig:\ref{fig:blockDistributionWithTask}).
This further supports the idea that using neuro-correlates does not necessarily create negative synergy as discussed in Results and Discussions concerning multiple parameter optimization.

\begin{figure}
\begin{center}
  \includegraphics[width=0.5\columnwidth]{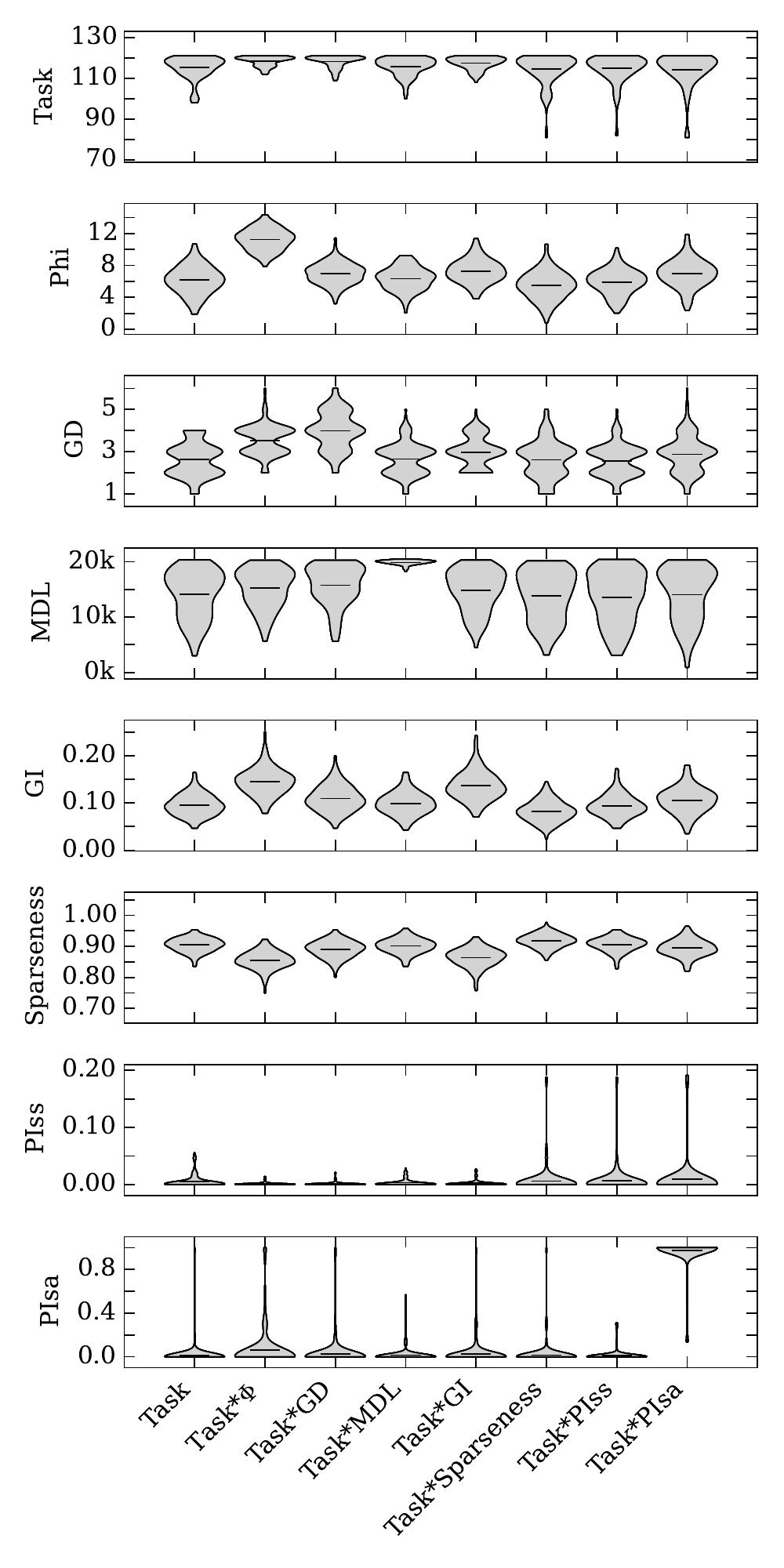} 
\end{center}
\caption{Comparison of different optimization methods on each other at the end of evolution of the random number generation task. The top row shows the performance distribution (gray) and average performance given the different selection regimes (MDL abbreviates minimum-description-length, which is the potential brain size based on the length of the encoding genome, GI abbreviates Gamma Index, GD graph diameter, and $PI_{\text{ss}}$ and $PI_{\text{sa}}$ are predictive information sensor $t$ to sensor $t+1$, and sensor $t$ to actuator $t+1$, respectively). Subsequent rows show the effect of the different selection regimes ($x$ axis) on the respective neuro-correlates.}
\label{fig:randomDistribution}
\end{figure}

In the RNG environment $R$ cannot be used because there is no environment about which the agent could build representations. For other neuro-correlates we observe the same trend in the RNG task as in the block catching task (see Figure:\ref{fig:randomDistribution}). Selecting for a particular neuro-correlate increases its measure over evolutionary time, more so than the increase found without explicit selection. The exception to this is Gamma Index and sparseness and $\Phi_{\text{atomic}}$. All other neuro-correlates seem to have no additional effect on each other.

\newpage
\subsection{Neuro-correlate interactions}

We showed that selection can be augmented by using specific neuro-correlates while others do not help. Is that because selecting for the neuro-correlates themselves already provides an advantage for performance? One might think that for example maximizing knowledge about the world ($R$) requires the agent to move in such a fashion that performance enhances automatically. Therefore we repeated the above described experiment, but instead of augmenting performance with a neuro-correlate, this time selection was performed on the neuro-correlates alone.
\begin{figure}[t]
\begin{center}
\includegraphics[width=0.4\columnwidth]{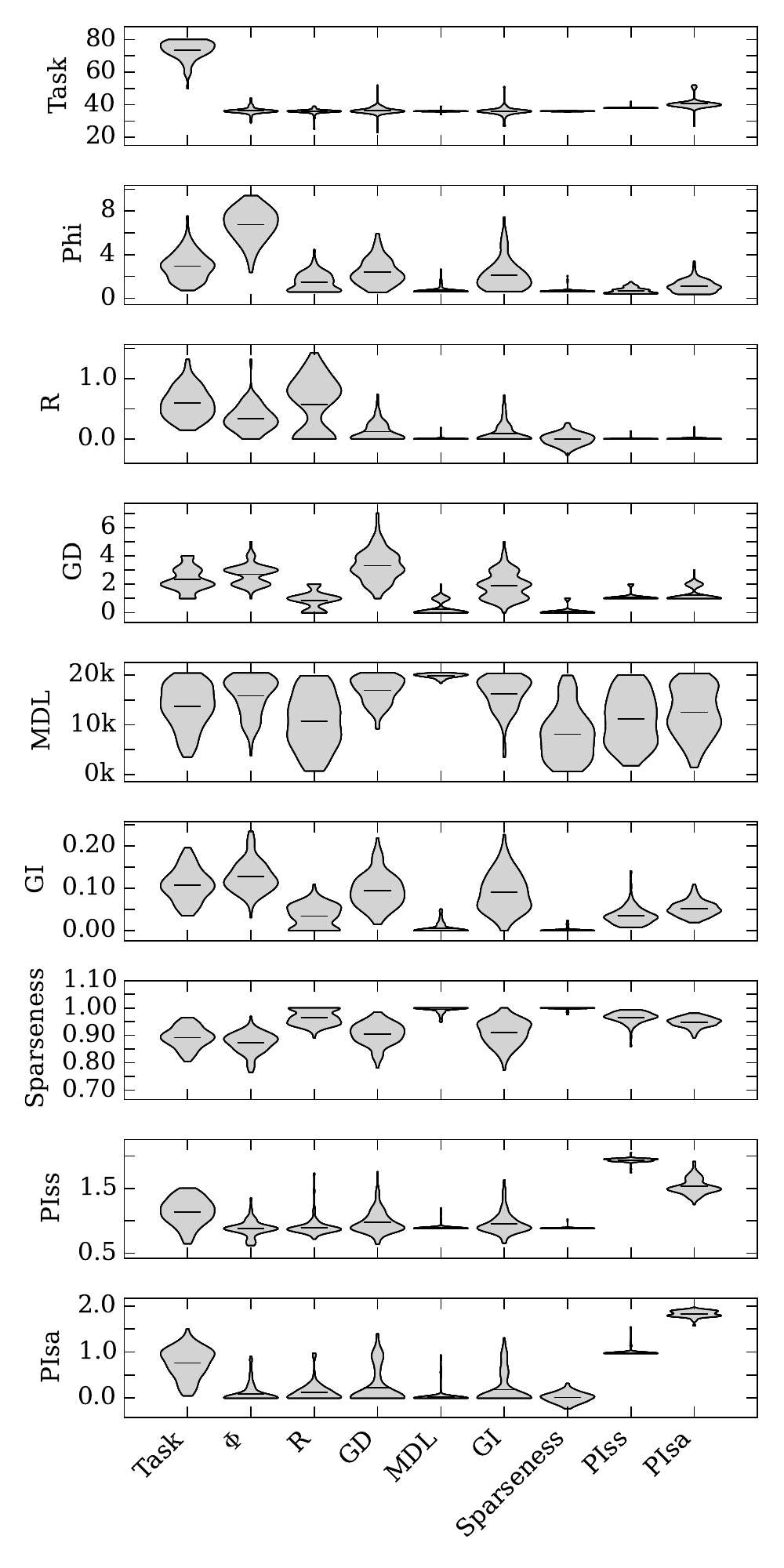} 
\includegraphics[width=0.4\columnwidth]{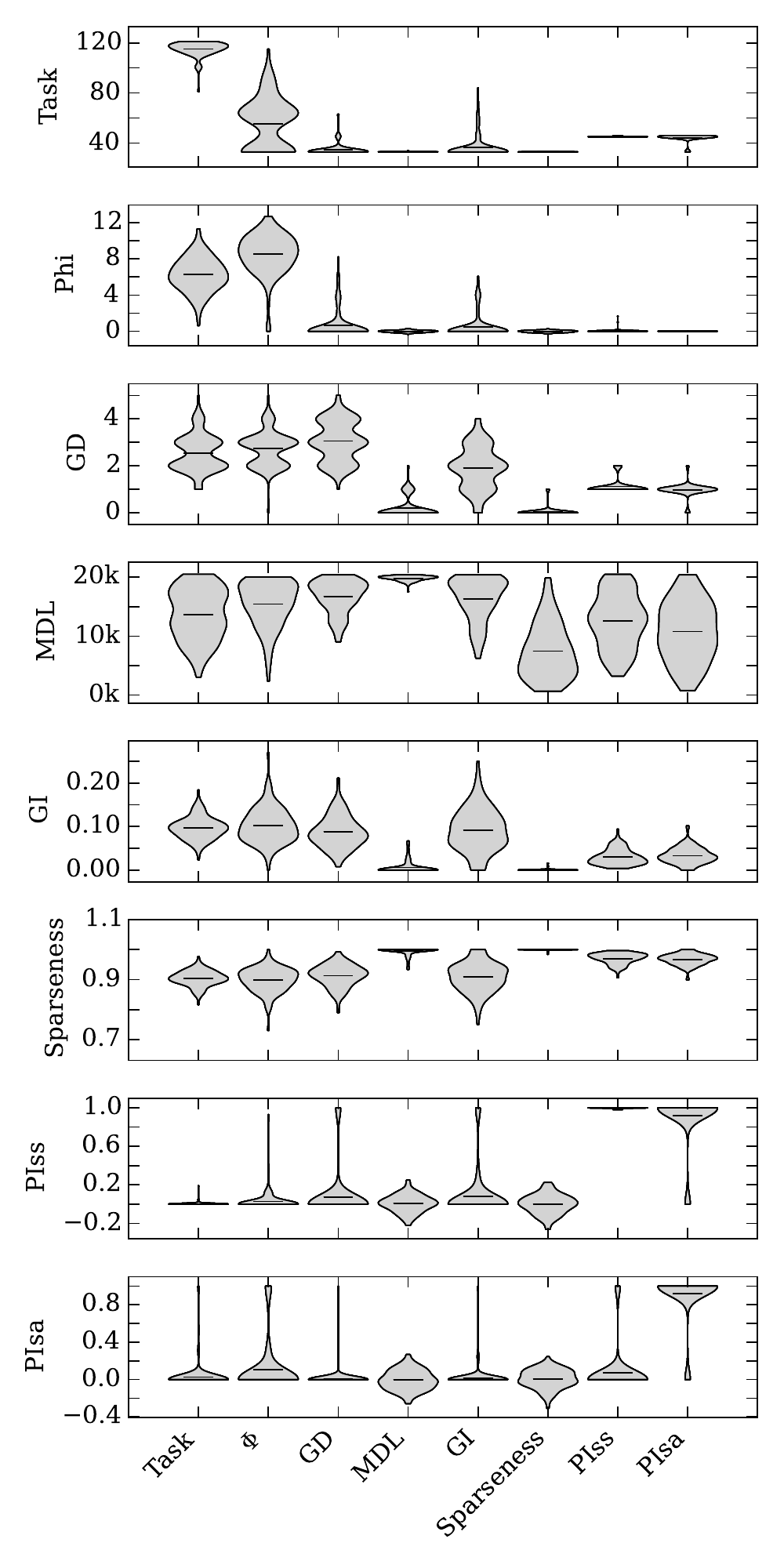} 
\end{center}
\caption{Effect on different neuro-correlates and performance when selecting only on a single neuro-correlate. The left panel is about evolution in the block catching environment, and the panel on the right is about the RNG task.}
\label{fig:singleDistribution}
\end{figure}

We find that none of the neuro-correlates affect performance substantially in the context of the RNG task (see the top rows in Figure:\ref{fig:singleDistribution}) except for $\Phi_{\text{atomic}}$. The effect on $\Phi_{\text{atomic}}$ on the generation of random numbers is measurable, but very low. We assume that in order to generate non-zero $\Phi_{\text{atomic}}$ information must flow through the system. Because there is no control function the ``information'' is literally random (entropy), which is what the RNG environment seems to be selecting.

As expected, selecting for a single neuro-correlate increases its value in both environments (see the diagonal for both environments in Figure:\ref{fig:singleDistribution}). However, we also find that many neuro-correlates affect each other positively and negatively, and the effect is similar in both environments. Some of these interactions are very intuitive. All measures that benefit from more connections, for example, cause the minimum-description-length to increase, whereas sparseness causes the minimum-description-length to shrink. Similarly, $\Phi_{\text{atomic}}$ and $R$ have at least some positive effect on each other, and we conjecture that having the ability to form representations even though they might not be used to improve performance still requires the ability to integrate information.

However, $PI_{\text{ss}}$ positively affects $PI_{\text{sa}}$ in the block catching environment and has no effect in the RNG environment, while $PI_{\text{sa}}$ has a positive effect on $PI_{\text{ss}}$ in both environments (compare the bottom right of each Figure:\ref{fig:singleDistribution}). To our knowledge the relation between the two Predictive Information measures has not been studied, and we are unable to provide any additional insight into this phenomenon.

\section{Conclusions}

We have tested whether eight different network- and information-theoretic neuro-correlates can be used to improve the performance of a simple genetic algorithm. We found that $\Phi_{\text{atomic}}$, $R$, graph diameter, and density (Gamma Index) each generally improve the performance of a GA in two environments tested, (one environment for the case of $R$). Sparseness does not improve performance as much as density does, suggesting sparseness is not generally beneficial for GAs in this domain. Thus, sparseness should only be used if its application has been shown to be beneficial for the problem domain in question. The two forms of predictive information measures ($PI_{\text{ss}}$ and $PI_{\text{sa}}$) had a negative effect in both environments on finding optimal performers (see Figure:\ref{fig:RatioPerfect}), and thus should not be used to augment selection in these kinds of environments.

Because the value of each neuro-correlate is simply multiplied by performance, the computational overhead is bound by the complexity of each measure. Typically, $R$ and $\Phi_{\text{atomic}}$ measures are computationally intensive and must be repeated for every agent in the population. This is a significant overhead, especially for increases in agent lifetime or brain size. However, this study shows graph diameter and gamma index measures to be computationally inexpensive and thus preferable, with preference between the two for graph diameter.
While the GA used here benefited from augmenting selection with neural correlates, we studied only two environments, and it is likely other environments might respond differently. Another possible extension of this work is to investigate other neuro-correlates, or even more topologically based neuro-correlates, or perhaps other algorithms such as novelty search~\cite{Lehman2008}.

\section*{Acknowledgments}
This work was supported in part by the National Science Foundation's BEACON Center for the Study of Evolution in Action under Cooperative Agreement DBI-0939454. We wish to acknowledge the support of the Michigan State University High Performance Computing Center and the Institute for Cyber Enabled Research (iCER).

\section*{Author Contributions}
AH wrote the core computational evolution code. JS wrote the analysis code. JS performed all computational analysis. All authors conceived the experiments, evaluated the results, and wrote the manuscript.

%



\subsection*{\noindent Supplementary Material}

\setcounter{figure}{0}

\makeatletter 
\renewcommand{\thefigure}{S\@arabic\c@figure}
\makeatother

\begin{figure}[H]
\begin{center}
    \textbf{Evolutionary History Cross Interaction for Block Catching}\par\medskip
  \includegraphics[width=\columnwidth]{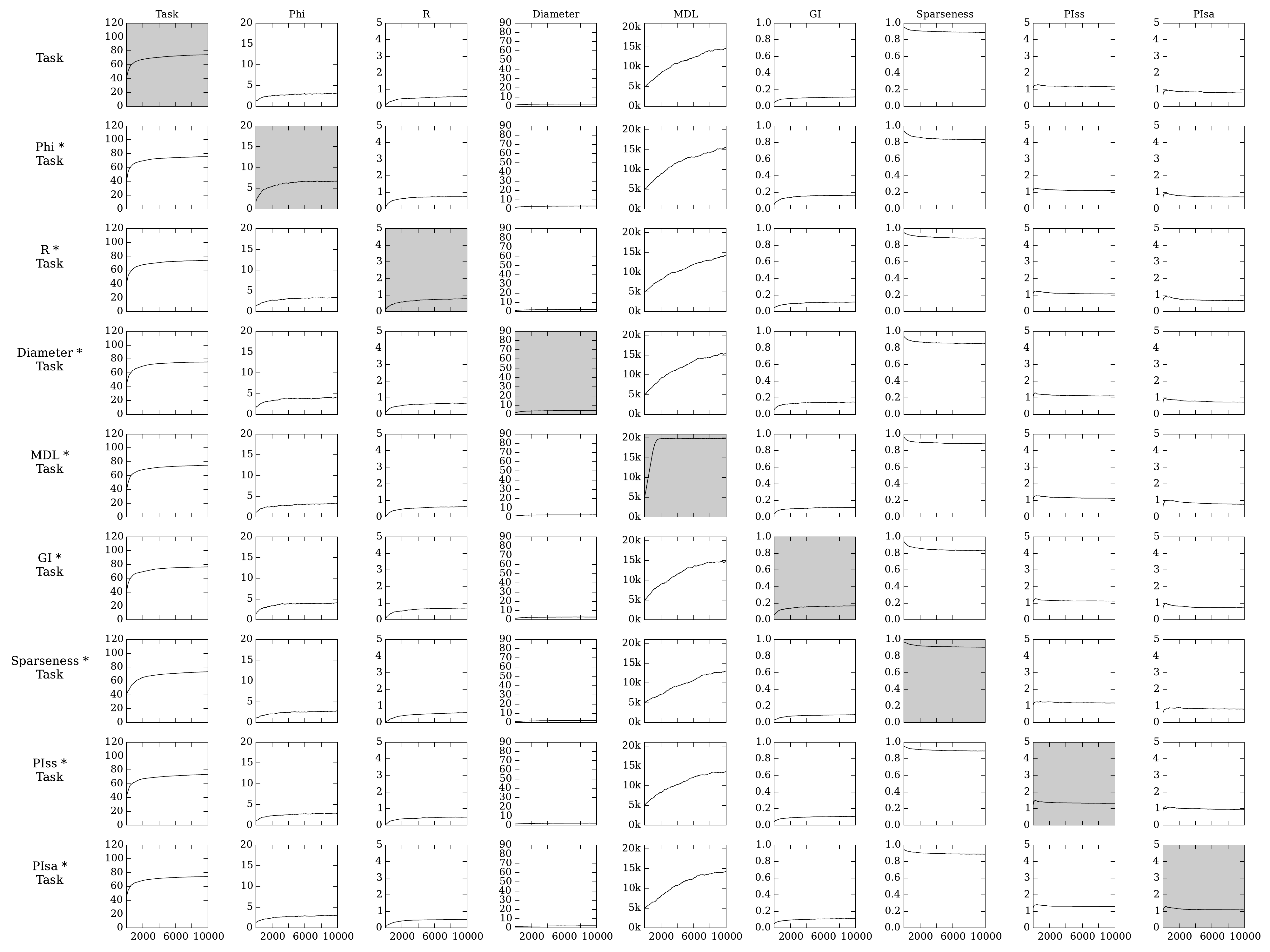} 
\end{center}
\caption{Selection for both task and a cognitive trait and the effect on each cognitive trait. Task (Block Catching Task) was selected for, in adition to each $y$ axis trait. the $x$ axis is the trait shown over evolutionary time for the $y$ axis treatment. The populations were allowed to evolve for $10,000$ generations and the results averaged over $128$ replicates.}
\label{fig:crossComparisonBlockDouble}
\end{figure}

\begin{figure}[H]
\begin{center}
    \textbf{Evolutionary History Cross Interaction for Random Number Generation}\par\medskip
  \includegraphics[width=\columnwidth]{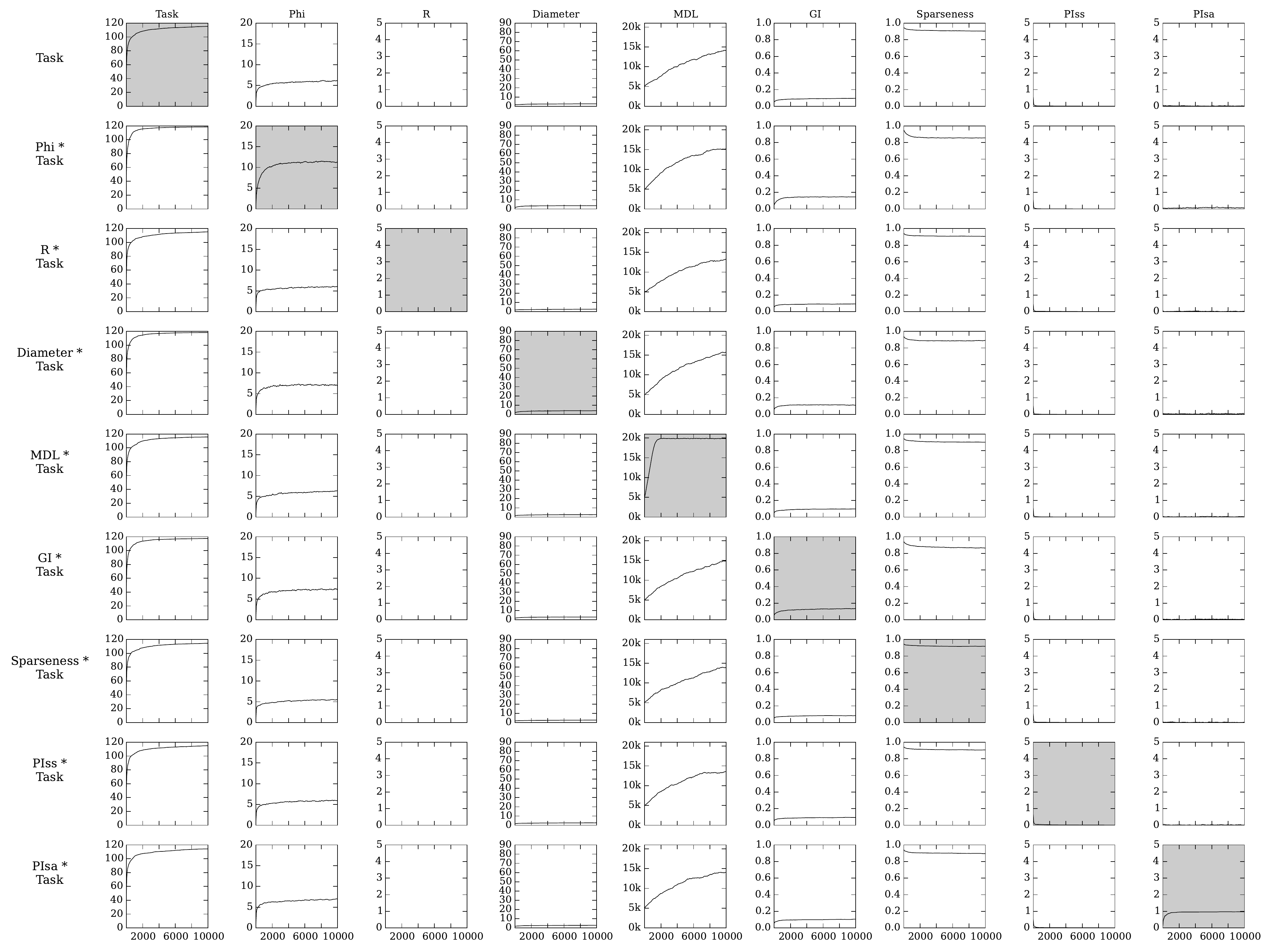} 
\end{center}
\caption{Selection for both task and a cognitive trait and the effect on each cognitive trait. Task (Random Number Generation Task) was selected for, in adition to each $y$ axis trait. the $x$ axis is the trait shown over evolutionary time for the $y$ axis treatment. The populations were allowed to evolve for $10,000$ generations and the results averaged over $128$ replicates.}
\label{fig:crossComparisonRandom}
\end{figure}

\begin{figure}[H]
\begin{center}
    \textbf{Evolutionary History Single Cross Interaction for Block Catching Task}\par\medskip
  \includegraphics[width=\columnwidth]{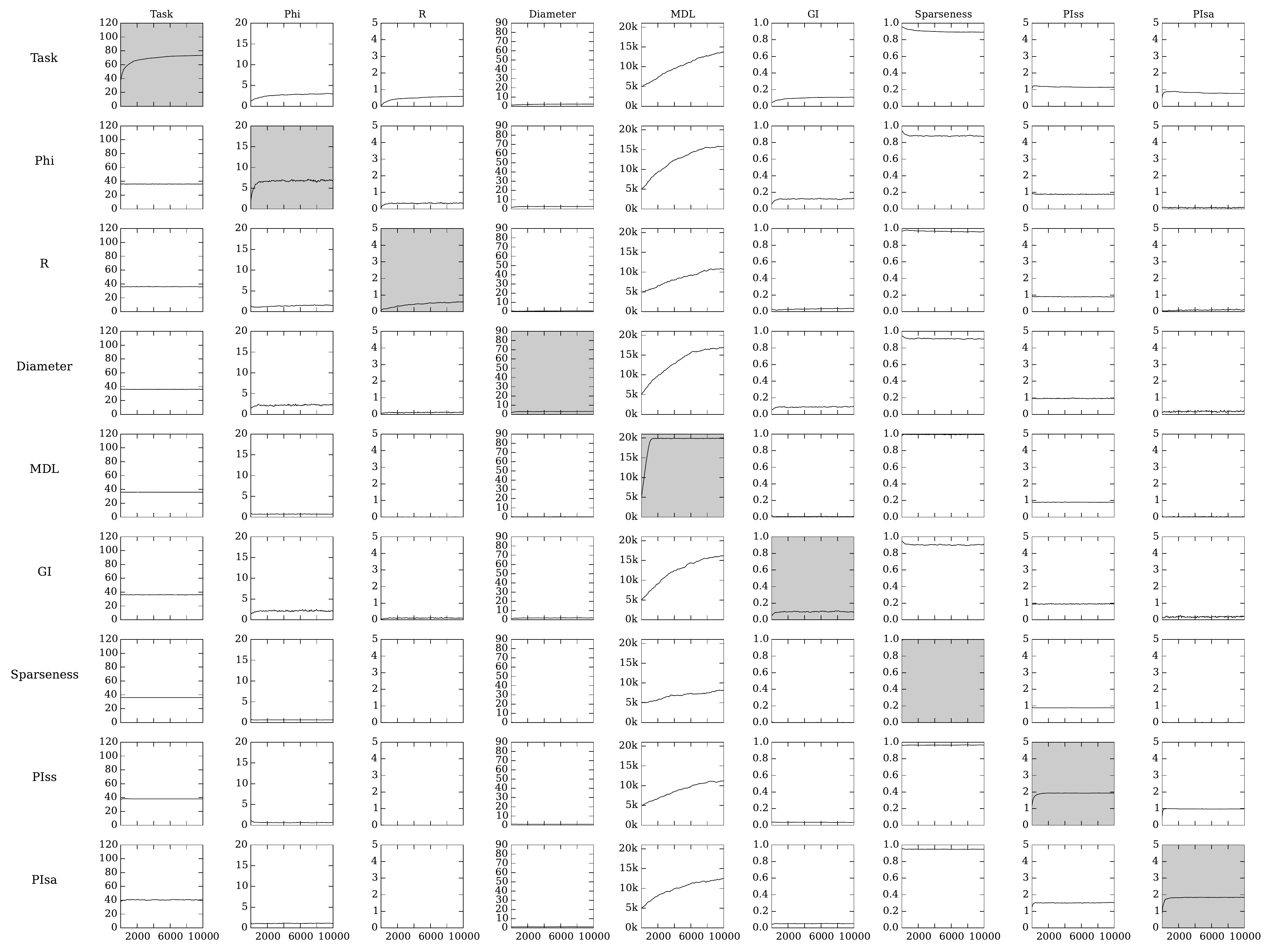} 
\end{center}
\caption{Selection for only task or a single cognitive trait and the effect on task and each cognitive trait. The $y$ axis presents for what was selected and the $x$ axis represents the affected trait shown over evolutionary time for the $y$ axis treatment. The populations were allowed to evolve for $10,000$ generations and the results averaged over $128$ replicates.}
\label{fig:crossComparisonRandomSingle}
\end{figure}

\end{document}